\documentstyle[epsfig]{aipproc}

\begin{document}

\title{A Dynamical $\eta'$--Mass from an Infrared Enhanced Gluon
Exchange}

\author{Lorenz von Smekal$^*$, Almut Mecke$^\dagger $ 
and Reinhard Alkofer$^\dagger $} 
\address{$^*$Physics Division, Argonne National Laboratory, Argonne, IL,
60439\\ $^\dagger$Institut f\"ur Theoretische Physik, Universit\"at
T\"ubingen, 72076 T\"ubingen, Germany\\[10pt]
\rm\footnotesize\noindent  ANL-PHY-8768-TH-97   \hfill UNITU-THEP-14/1997
\hfill  hep-ph/9707210\\
\it Talk presented at the 6th Conference on the Intersections of Particle and
Nuclear Physics, Big Sky, Montana, 5/27 - 6/2, 1997.}
\maketitle
\vskip -1cm

\begin{abstract} 
The pseudo--scalar flavor--singlet meson mixes with two gluons. A dimensional
argument by Kogut and Susskind shows that this can screen the Goldstone pole
of the chiral limit in this channel, if the gluon correlations are infrared
enhanced. Using a gluon propagator as singular as $\sigma/k^4$ for $k^2 \to
0$ we relate the screening mass to the string tension $\sigma$. In the
Witten--Veneziano action to describe the $\eta$--$\eta'$ mixing this relation
yields masses of about 810MeV for the $\eta'$, 430MeV for the $\eta$ and a
mixing angle of about $-30^\circ$ from the phenomenological value $\sigma
\approx  0.18$GeV$^2$. The very weak temperature dependence of the string
tension should make this mechanism experimentally distinguishable from
exponentially temperature dependent instanton model predictions.  
\end{abstract}

\vskip -.2cm
More than twenty years ago Kogut and Susskind pointed out that for
dimensional reasons a non--vanishing contribution to the mass of the
pseudo--scalar flavor--singlet meson in the chiral limit can result from
its mixing with two non--perturbatively infrared enhanced gluons
corresponding to a momentum space propagator $D(k) \sim \sigma/k^4$ for $k^2 
\to 0$ \cite{KS74}. Such infrared enhanced gluon correlations are known to
lead to an area law in analogy to the Schwinger model in two dimensions. 
The identification of the string tension $\sigma$ shows that effects due to
infrared enhanced gluons can be expected to be complementary to instanton
models. In particular, a description of the $\eta$--$\eta'$ mixing driven by
the string tension \cite{MSA97}, provides an interesting alternative to the
standard solution of the U$_A(1)$ problem by instantons. 

Phenomenologically, this mixing is described by the  $\eta_8 - \eta_0$  mass
matrix \cite{WV79}, 
\begin{eqnarray}
 \frac{1}{2}\ (\eta_8\ \ \  \eta_0)
 \left(
 \begin{array}{cc}
  \frac{4}{3}m_K^2-\frac{1}{3}m_\pi^2&\frac{2}{3}\sqrt{2}(m_\pi^2-m_K^2)\\
   & \\
  \frac{2}{3}\sqrt{2}(m_\pi^2-m_K^2) &\frac{2}{3}m_K^2+\frac{1}{3}m_\pi^2
                                       + \frac{2N_f}{f_0^2} \chi^2
 \end{array}
 \right)
 \left(
 \begin{array}{c}
  \eta_8\\
   \\
  \eta_0
 \end{array}
 \right)
\end{eqnarray}
where the screening mass in the flavor--singlet component, $m_0^2 := 2N_f
\chi^2/f_0^2 $, is given by a non--vanishing topological susceptibility,
\begin{equation} 
\chi^2 := \frac{g^2}{(32\pi^2)^2} \int d^4x \, \langle \widetilde GG(x) \,
\widetilde GG(0) \rangle \quad \hbox{with}  
\end{equation} 
\begin{displaymath}
 \widetilde GG  =  \epsilon^{\mu\nu\rho\sigma} 2\partial_\mu \, {\rm tr}
(A_\nu  \partial_\rho A_\sigma -ig \frac 2 3 A_\nu A_\rho A_\sigma ) \; .
\end{displaymath}
In the Instanton Liquid Model the topological susceptibility, given by the
density of instantons, is $\chi^2 \approx 1 \hbox{fm}^{-4}$, and the mass
eigenvalues are $m_\eta \approx 530$MeV, $m_{\eta'} \approx 1170$MeV 
together with a mixing angle of $\theta \approx -11.5^\circ$ \cite{ANVZ89}.    

\begin{figure}[t]
\centerline{\epsfig{file=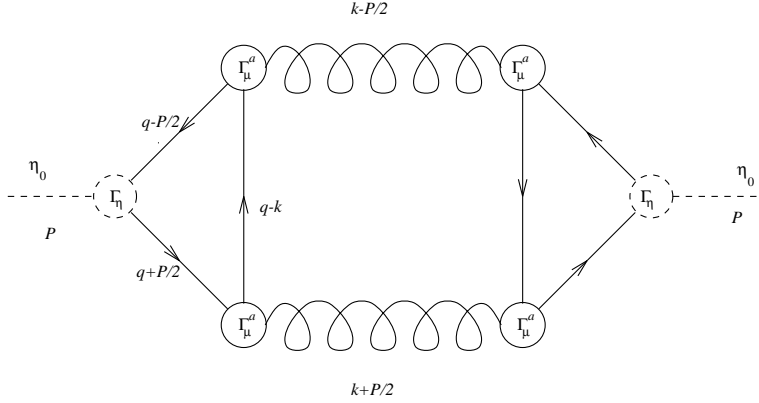,height=5cm,width=10cm}}
\vspace{5pt}
\caption{The diamond diagram $\Pi (P^2)$, a factor 2 arises from crossed
gluon exchange.}  
\label{diamond}
\end{figure}

Here, we concentrate on the mixing of the flavor--singlet pseudo--scalar with
two uncorrelated gluons. According to the Kogut--Susskind argument, for
infrared enhanced gluons $\sim \sigma/k^4$, the corresponding diagram, see
fig. \ref{diamond}, can contribute to the topological susceptibility for
the meson momentum $P \to 0$. To explore this conjecture and its quantitative
consequences, we use the following model interaction for quarks in the Landau
gauge, 
\begin{equation} 
g^2 D_{\mu\nu}(k) = P_{\mu\nu} (k) \, \left( \frac{8\pi \sigma}{k^4} \, +\,
\frac{16\pi^2/9}{k^2 \ln(e+k^2/\Lambda^2)} \right) \; .
\label{int}\end{equation}
The second term, subdominant in the infrared, was added to simulate the
effect of the leading logarithmic contribution of perturbative QCD for $N_f =
3$. Strictly speaking, a quark interaction of the form (\ref{int}) cannot
arise from gluons alone in Landau gauge, since the product $g^2 D_{\mu\nu}$
is not renormalization group invariant for any finite number of flavors or
colors. Even though this is assumed in the {\it Abelian} approximation, ghost
contributions do implicitly enter in the RG invariant interaction (by the
dressing of the quark--gluon vertex function). In fact, three quite different
approaches to the pure gauge theory are available at present to
suggest that the strong infrared enhancement of the interaction might be
generated by ghost contributions in Landau gauge \cite{LG}.   

From the axial anomaly, the quark triangle $\Gamma^{ab}_{\mu\nu}$ in
fig. \ref{diamond} has the limit,\\
$ P \to 0 \, , k^2 = 0 \; : \qquad  \Gamma^{ab}_{\mu\nu} \, \to \, \delta^{ab} \epsilon_{\mu\nu\rho\sigma} \,
k^\rho P^\sigma \, \sqrt{N_f} f_0^{-1} g^2/(8\pi^2) $\\
This model independent form, determining the coupling of two gluons to the
pseudo--scalar flavor--singlet bound state in the infrared, is particularly
suited for the present calculation, since the contribution to $\chi^2 $ is
obtained from $P\to 0$, and since the gluon interaction (\ref{int}) weights
the integrand so strongly in the infrared ($\sim \sigma/(k\pm P/2)^4$). With
this, all contributions containing ultraviolet dominant terms of the
interaction (\ref{int}) vanish for $P\to 0$, and we obtain \cite{MSA97},
\begin{equation}
m_0^2 \, = \, \lim_{P^2\to 0} \Pi(P^2) \, =\, \frac{2 N_f}{f_0^2} \, \chi^2 \, =\, \frac{3N_f}{f_0^2} \,
\frac{\sigma}{\pi^4} \; .
\end{equation} 
The phenomenological string tension $\sigma = 0.18$GeV$^2$ and $f_0\approx
f_\pi =93$MeV thus yield $m_0^2\approx 0.346$GeV$^2$, and the physical mass
eigenstates are, $m_{\eta^\prime} \approx 810$MeV and $m_{\eta} \approx
430$MeV, with a corresponding mixing angle $\theta \approx -30^\circ$. 

Using free constituent quarks of a mass of about $300$MeV in the triangle to
suppress spurious ultraviolet contributions, 
from $f_0^2 \simeq  f_\pi^2 ( 1 + \Pi'(P^2) |_{P^2 \to 0} )$ with
$\Lambda\approx 500 $MeV in (\ref{int}), we obtain an additional contribution
to the decay constant of the flavor--singlet of about 30\% as compared to the
pion \cite{MSA97}.   

As these values are reasonably close to experiment, we conclude that the
$U_A(1)$--anomaly might be encoded in the infrared behavior of QCD Green's
functions. Whether the Kogut--Susskind mechanism or the instanton based
solution to the U$_A(1)$ problem is realized in nature, can be assessed from
their respective temperature dependences. If the origin of the $\eta^\prime$
mass is predominantly due to instantons, the $\eta - \eta^\prime$ mixing
angle is expected to vary exponentially with temperature, leading to a
significant change of $\eta$ and $\eta^\prime$ production rates in
relativistic heavy ion collisions \cite{AAR89}. On the other hand, lattice
calculations indicate that the string tension is almost temperature
independent up to the deconfinement transition. This offers the possibility
to study the physics of the U$_A(1)$ anomaly experimentally. 


We thank T.-S.~H.~Lee, H.~Reinhardt and C.~D.~Roberts
for helpful discussions. RA gratefully acknowledges the hospitality 
of the Physics Division at ANL. 
This work was supported by the DFG under contract
Al 279/3-1 and the US-DOE, Nuclear Physics Division, contract \#\
W-31-109-ENG-38.

\end{document}